\documentclass[prb,twocolumn,showpacs]{revtex4-1}
\usepackage{epsfig}
\usepackage{graphicx}
\usepackage{amsfonts}
\usepackage[figuresright]{rotating}
\usepackage{amssymb}
\usepackage{amsmath}
\usepackage{psfrag}
\usepackage{bm}
\usepackage[colorlinks,linkcolor=blue,anchorcolor=blue,citecolor=blue,urlcolor=blue]{hyperref}

\def\avg#1{\langle#1\rangle}

\def\be{\begin{equation}} \def\ee{\end{equation}}
\def\bea{\begin{eqnarray}} \def\eea{\end{eqnarray}}

\def\nn{\nonumber}

\newcommand{\br}{\textbf{r}}

\begin{document}

\title{Stability of the Nagaoka-type Ferromagnetic State in a $t_{2g}$ Orbital System on a Cubic Lattice}
\author{Eric Bobrow}
\author{Yi Li}
\address{Department of Physics and Astronomy, The Johns
Hopkins University, Baltimore, MD 21218, USA}
\date{April 6, 2018}

\begin{abstract}
We generalize the previous exact results of the Nagaoka-type itinerant ferromagnetic states in a three dimensional $t_{2g}$-orbital system 
to allow for multiple holes. 
The system is a simple cubic lattice with each site possessing
$d_{xy}$, $d_{yz}$, and $d_{xz}$ orbitals, which allow two-dimensional hopping within each orbital plane.
In the strong coupling limit of $U\to \infty$, the orbital-generalized
Nagaoka ferromagnetic states are proved degenerate with the ground
state in the thermodynamic limit when the hole number per orbital
layer scales slower than $L^{\frac{1}{2}}$.
This result is valid for arbitrary values of the ferromagnetic Hund's coupling $J>0$ and inter-orbital repulsion $V\ge 0$.
The stability of the Nagaoka-type state at finite electron densities with respect to a single spin-flip is investigated.
These results provide helpful guidance for studying the mechanism
of itinerant ferromagnetism for the $t_{2g}$-orbital materials.
\end{abstract}
\maketitle


\section{Introduction}
\label{sect:intro}
Itinerant ferromagnetism, i.e., ferromagnetism of metallic states with Fermi surfaces, remains a challenging problem of condensed matter physics.
Since the Coulomb interaction is spin-independent, it cannot directly
give rise to electron spin polarizations at the classic level, and
the mechanism of itinerant ferromagnetism is hence fundamentally quantum mechanical.
As first described through Stoner's criterion \cite{Stoner1938},
itinerant ferromagnetism arises from the direct exchange interaction among
electrons with the same spin.
However, this criterion overlooks correlation effects among electrons
with opposite spins.
In fact, even in the presence of very strong interactions, electrons
can remain unpolarized and build up highly correlated
wavefunctions to reduce repulsive interactions.
Due to the intrinsic strong-correlation nature of the problem, it is usually difficult to obtain a precise description based on
perturbative approaches, and thus non-perturbative results
have played important roles in the study of itinerant ferromagnetism \cite{Lieb1962,Nagaoka1966,Lieb1989,Mielke1991,Mielke1991a,
Mielke1992,Tasaki1992,Mielke1993a,Tasaki2003,Lieb1995,Hirsch1989,
Tanaka2007,Zhang2015a}.
The previously known exact results of ferromagnetism are largely
classified into two categories: Nagaoka ferromagnetism
\cite{Nagaoka1966,Tasaki1989}
for single-band Hubbard systems and flat-band ferromagnetism
\cite{Tasaki1992,Mielke1992}.


A key result in the study of itinerant ferromagnetism is Nagaoka's
theorem for the Hubbard model,
which proves the ground state to be fully spin-polarized for exactly
one hole away from half-filling in the $U \rightarrow \infty$ limit\cite{Nagaoka1966}.
The underlying physics is that the fully polarized state maximally
facilitates the hole's coherent hopping to reduce the kinetic energy.
Tasaki \cite{Tasaki1989} simplified the proof of Nagaoka's
theorem by using the Perron-Frobenius theorem \cite{Lieb1962}.
However, this method typically breaks down for fermionic systems with multiple holes in two or higher dimensions because of the fermion
sign originating from the antisymmetry under exchange.
For the case of a single-band Hubbard model in the $U \rightarrow \infty$
limit, the Nagaoka-type ferromagnetic state with multiple holes was
proven to be degenerate with the ground state under the
following conditions\cite{Tian1991,Shen1993a}: The hole number $N_h \sim L^{\alpha}$ for $0 \le \alpha < 1$ on an $L\times L$ square lattice
in the limit of $L\to \infty$.
This stability was investigated using a squeeze theorem argument in
which a variational upper bound to the ground state energy was
shown to be equal to a lower bound for suitably low $N_h$
in the thermodynamic limit.
The variational trial state used was the fully spin-polarized
Nagaoka-type state.
The lower bound was given by the Gershgorin circle theorem from
linear algebra, which is explained in the Appendix.

In contrast to single-band Hubbard models, most itinerant ferromagnetic metals are orbital-active. In such systems, the multi-orbital structure together with Hund's interaction play an important role in the onset of itinerant ferromagnetism despite Hund's interaction being local and typically
only polarizing spins on the same site.
Recently, exact results of itinerant ferromagnetism in strongly
correlated multi-orbital systems were proven, providing a sufficient condition for ferromagnetism driven by Hund's coupling \cite{Li2014}.
Different from the Nagaoka theorem, an entire phase of
itinerant ferromagnetism is set up with a wide range of electron filling.
Furthermore, since such systems are free of the fermion sign-problem of
quantum Monte-Carlo simulations \cite{Li2014}, the Curie-Weiss
metal phase and critical scalings of the ferromagnetic phase
transitions have been simulated by quantum Monte-Carlo to high
numerical precision \cite{Xusl2015}.
However, the proof of the ground state ferromagnetism is based on
the Perron-Frobenius theorem, which requires one-dimensional bands
to avoid issues related to fermionic exchange.
In order to generalize itinerant ferromagnetism to orbital-active systems with a quasi-two dimensional band structure, a $t_{2g}$-orbital system
was previously studied \cite{LiFM2015} with the number of holes
restricted to exactly one away from half-filling in each orbital plane.
In this case, a multi-orbital generalization of the Nagaoka-like
ferromagnetic state was proven to be the unique ground state up to
spin degeneracy. Nevertheless, its stability has not been previously studied.

In this article,  we extend the multi-orbital Nagaoka-type ferromagnetic
states proven in Ref. [\onlinecite{LiFM2015}] to the multi-hole
case and investigate its stability.
The analyses are done in the presence of two- and three-dimensional
band structures in the strong correlation regime.
By generalizing the method used in Refs. [\onlinecite{Tian1991,Shen1993a}]
to include Hund's coupling and inter-orbital repulsion,
we show that in the strong interaction limit, the generalized Nagaoka
ferromagnetic state with the $t_{2g}$-orbitals remains degenerate
with the ground state in the multi-hole case in the thermodynamic
limit.
Although the hole density remains zero, the hole number can go to
infinity, scaling as a finite power of the system size.
An analysis of the stability of the generalized Nagaoka state against flipping
a single spin is also performed, which shows that the region
of instability shrinks in the presence of Hund's coupling.

The remainder of the article is organized as follows:
In Sect. \ref{sect:model}, we introduce the multi-orbital Hubbard
model for a three-dimensional (3D) $t_{2g}$-orbital system with a
two-dimensional (2D) band in a cubic lattice.
In Sect. \ref{sect:stability}, we analyze the stability of the
ferromagnetic ground state in the presence of multiple holes for the
multi-orbital model presented in Sect. \ref{sect:model}.
Then, by analyzing the change in energy due to a single
flipped spin\cite{Shastry1990}, a region in which
the fully polarized state is no longer a ground state is identified.
The Nagaoka type ferromagnetic state and its stability in the presence
of a 3D band structure is studied in Sect. \ref{sect:3Dbands}.
Conclusions and discussions are presented in Sect. \ref{sect:conclusion}.

\section{The $t_{2g}$ Orbital system with multi-orbital interactions}
\label{sect:model}
In this section, we present the band structure of the 3D $t_{2g}$
orbital system and the on-site multi-orbital Hubbard interactions.

The system to be studied is a 3D multi-orbital Hubbard model on an $L\times L\times L$
simple cubic lattice with $d_{xy}$, $d_{yz}$, and $d_{xz}$ orbitals
at each site with on-site multi-orbital interactions.
The Hamiltonian can be written as
\begin{equation}
H = H^K + H^U + H^V+ H^J,
\end{equation}
where $H^K$, $H^U$, $H^V$, and $H^J$ represent the kinetic energy, the
intra-orbital Hubbard interaction, the inter-orbital Hubbard interaction, and the inter-orbital Hund's coupling, respectively.
Since the $t_{2g}$ orbitals are planar, the hopping term for each orbital
is anisotropic in general. 
For the system with quasi-2D band structure studied here and in Sect. \ref{sect:stability}, transverse hopping perpendicular to the orbital plane is much weaker
than intra-plane hopping and hence will be neglected. A fully 3D band structure including transverse perpendicular hopping terms will be considered in Sect. \ref{sect:3Dbands}.
With intra-plane hopping neglected, the kinetic term corresponding to the $d_{xy}$ orbital takes the form
\begin{equation}
\begin{aligned}
H^K_{xy} &= t\sum\limits_{\textbf{r},\sigma} \big(d_{xy,\sigma}^\dagger(\textbf{r})d_{xy,\sigma}(\textbf{r}+\hat{x}) \\
&+ d_{xy,\sigma}^\dagger(\textbf{r})d_{xy,\sigma}(\textbf{r}+\hat{y}) + h.c.\big),
\end{aligned}
\label{eq:kin}
\end{equation}
where the lattice constant is taken to be $1$ and $t$ is the hopping integral.
The hopping terms for the other orbital planes have the same form with the directional indices replaced as necessary, and the full kinetic term $H^K$
is a sum of hopping terms for the $d_{xy}$, $d_{yz}$, and $d_{xz}$ orbital planes.

For a simple cubic lattice with negligible transverse hopping, Eq. \eqref{eq:kin} and its $yz$ and $xz$
counterparts constitute all nearest neighbor
hoppings allowed by symmetry.
Different orbitals do not mix at this level due to the cubic symmetry of the system, which can be seen as follows.
Without loss of generality, consider an $x$-bond between the sites
$\textbf{r}$ and $\textbf{r}+\hat{x}$.
Since this bond is invariant under reflections with respect to both the $xy$-plane and the $xz$-plane, hopping along this bond should respect these symmetries. The $d_{xy}$ orbital is even while $d_{xz}$ and $d_{yz}$ orbitals are odd under the former reflection. Thus, $d_{xy}$ does not mix with either $d_{xz}$ or $d_{yz}$ through this hopping. Furthermore, $d_{xz}$ is even while $d_{yz}$ is odd under the latter reflection, and thus they do not mix either.

The on-site multi-orbital Hubbard interactions consist of intra-orbital and inter-orbital terms.
The intra-orbital interaction $H^U$ is expressed as
\bea
 H^U = U \sum\limits_{\textbf{r}, a} n_{a, \uparrow}(\textbf{r})
  n_{a, \downarrow}(\textbf{r}),
\eea
where $a$ is the orbital index and $n_{a,\sigma}({\bf r}) = d^\dagger_{a,\sigma}({\bf r}) d_{a,\sigma}({\bf r})$.
The inter-orbital interaction takes the form
\begin{equation}
  H^V = V\sum\limits_{\textbf{r},a> b}
  (1-n_a(\textbf{r})) (1-n_b(\textbf{r})),
\end{equation}
where $n_a(\textbf{r}) = n_{a,\uparrow}(\textbf{r}) + n_{a,\downarrow}(\textbf{r})$.
$H^V$ is expressed in terms of the hole number occupation, which is equivalent
to the corresponding electron number form up to an overall constant.
Since we will explore the stability of the Nagaoka state in which
nearly every orbital on every site is filled, the hole representation
will be more convenient.

The final interaction in the model, the on-site inter-orbital Hund's coupling, reads
\begin{equation}
\begin{aligned}
  H^J &= -J \sum\limits_{\textbf{r}, a> b} \left(\textbf{S}_a(\textbf{r})
  \cdot \textbf{S}_b(\textbf{r}) -\frac{1}{4}
   n_a(\textbf{r})n_b(\textbf{r})\right).
\end{aligned}
\end{equation}
For any two orbitals $a$ and $b$ on site $\textbf{r}$,
the energy from the Hund's coupling is non-negative if $J>0$. The energy contribution is $J$
if both orbitals are filled and form a spin
singlet and zero otherwise.

Below, we will consider the limit of  $U \rightarrow \infty$, in
which no individual orbital can hold two electrons.
Instead, individual sites can hold up to three electrons,
all in different orbitals, with their interaction determined by
$H^V$ and $H^J$.
We also only consider the case of $V\ge 0$, i.e., with
repulsive inter-orbit interaction, and $J>0$, i.e.,
ferromagnetic Hund's coupling.

\section{Stability of the generalized Nagaoka-like state}
\label{sect:stability}
In this section, we investigate the stability of the
generalized Nagaoka state in the 3D $t_{2g}$ orbital systems
with quasi-2D band structure.
Such a state with a single hole per orbital layer was proven
previously in Ref. [\onlinecite{LiFM2015}].

Since the hopping term $H^K$ only allows holes to hop within orbital
planes, the number of holes in each orbital plane is conserved.
For simplicity, assume that each orbital plane has the same number $n_h$ of holes.
Since there are $L$ layers for each of the three orbital plane
directions, the total number of holes is then $N_h = 3Ln_h$.
Note that in what follows, the number of holes will always refer
to the number of holes above the half-filled background, or the number of electrons below half filling.

As in Ref. [\onlinecite{LiFM2015}], conservation of hole number in each orbital
plane allows the overall Hilbert space to be written as a tensor product
of Hilbert spaces for each layer in each orbital plane direction.
For the $l$-th layer of orbital type $a$, we define a reference state
$|R_{a,l,\uparrow}\rangle$ in which each orbital is filled with a spin-$\uparrow$
electron.
$|R_{a,l,\uparrow}\rangle$ is equivalent to  the state with all single-particle momentum states $\mathbf{k}$
in the 2D Brillouin zone fully filled.
Now we add $n_h$ holes by removing $n_h$ electrons
one by one from the highest filled single particle state.
The resulting many-body state, a Slater determinant state with all
electron spins up, is expressed as
\bea
|h_{a,l,\uparrow}\rangle=\prod_{i=1}^{n_h} d_{a,l}(\mathbf{k}_i)
|R_{l,a,\uparrow}\rangle,
\label{eq:slater}
\eea
where $d_{a,l}(\mathbf{k}_i)=\frac{1}{L}\sum_{\mathbf r} d_{a,l}(\mathbf r)
e^{i\mathbf k_i  \mathbf r}$ and $\mathbf{k}_i$ and $\mathbf {r}$ represent
2D momentum and lattice vectors, respectively. The momenta take values $\mathbf{k}_i=(\frac{2m_1\pi}{L},\frac{2m_2\pi}{L})$
with $m_{1,2}$ integers.
We consider the limit of $n_h/L^2\to 0$, where the single particle spectrum becomes parabolic.
From the sign convention of $t$ in Eq. \eqref{eq:kin},  $m_1$ and $m_2$
start from $(0,0)$ and take values in 
ascending order of $m_1^2+m_2^2$.

\subsection{Estimation of the upper bound}
In order to show the existence of a fully spin-polarized ground state,
consider a trial state
\begin{equation}
  \begin{aligned}
    |\psi_t\rangle=\bigotimes_{l=1}^L |h_{xy,l,\uparrow}\rangle
    \otimes |h_{yz,l,\uparrow}\rangle \otimes |h_{xz,l,\uparrow}\rangle,
\end{aligned}
\end{equation}
where the constraint of $n_h$ holes per layer is enforced by the form
of the basis states.
This is a fully spin-polarized state with the maximum
spin $S = S_z=\frac{N_s}{2}$ for $N_s$ spins.
Since the Hamiltonian possesses SU(2) symmetry, applying
$S_{T,-}=S_{T,x}-iS_{T,y}$ successively on $|\psi_t\rangle$ where
$\vec S_T$ is the total spin operator produces
a $2N_s+1$ SU(2) multiplet.

An upper bound on the ground state energy $E_g$ is derived by
evaluating the energy expectation value $E_T$ of the Nagaoka-like trial
state $|\psi_t\rangle$
\bea
E_T= E_{K}+E_{U}+E_{J}+E_{V},
\eea
where $E_{K}=\avg{\psi_t|H_K|\psi_t}/\avg{\psi_t|\psi_t}$ and
expressions for $E_U$, $E_J$, and $E_V$ can be defined
similarly.
$E_U=0$ since every individual orbital is at most singly occupied in the $U \to \infty$ limit.
Since $|\psi_t\rangle$ describes a fully polarized state,
any two electrons form a spin triplet and $E_J=0$ as well.
$E_V$ can be evaluated easily by noting that the hole distributions on different layers are uncorrelated for this trial state, and thus
\bea
E_V &=& V\sum_{r,a>b} \frac{\avg{\psi_t | 1-n_a(\mathbf{r})| \psi_t}\avg{ \psi_t | 1-n_b(\mathbf{r}) | \psi_t} }{\avg{\psi_t| \psi_t}^2} \nn \\
&=& 3\frac{n^2_h}{L} V.
\eea
The upper bound on the kinetic energy can be estimated as follows.
$E_K$ is the sum of the kinetic energies of each layer. Up to a constant, the dispersion for each band can be rewritten in terms of hole number occupation as
\bea
H^K_{xy} = -4t\sum_{\textbf{k}, \sigma}\left(1 - \frac{\mathbf{k}^2}{4}\right)(1-n_{xy, \sigma}(\mathbf{k}))
\eea
with parabolic dispersion near $k = 0$ in the $n_h/L^2 \to 0$ limit. Expressions for $H^K_{yz}$ and $H^K_{zx}$ follow by changing the indices. Since the trial state corresponds to removing $n_h$ electrons from the band maximum of the fully filled Brillouin zone, or equivalently adding $n_h$ holes to the band minimum in the hole description, the kinetic energy for a single layer can be estimated as
\bea
-4 n_h t + t L^2 \int_0^{k_0} \frac{k^3 dk  }{2\pi}
= -4n_h t + t O( \frac{n_h^2}{L^2}),
\eea
where $(L/2\pi )^2 \pi k_0^2\approx n_h$.
Summing over all $3L$ layers gives
\bea
E_K = -12 n_h Lt + t O(\frac{n_h^2}{L}).
\eea
Including the $E_V$ contribution, the total trial state energy serves as an upper bound on the ground state energy $E_g$ of
\bea
E_g \leq -12 n_h Lt + t O(\frac{n_h^2}{L}) + 3 V \frac{n_h^2}{L}.
\eea

\subsection{Estimation of the lower bound }

Since both $H^J$ and $H^V$ are non-negative operators, their lower bounds
are zero.
Thus, a lower bound on $H^K$ is also a lower bound on $E_g$.
Since $H^K$ is the sum of the kinetic energies of each layer,
the sum of the lower bounds on the kinetic energy of each layer
is also a lower bound on $E_g$.
The lower bound on the kinetic energy of each layer is simply $-4n_ht$, which
has been worked out \cite{Tian1991,Shen1993a} by applying the Gershgorin
circle theorem and considering a configuration where each hole has
no neighboring holes.
A brief review of this result is provided in the Appendix.
Summing over each layer, we arrive at the lower bound on the ground
state energy
\bea
-12n_h Lt \le E_g.
\eea
Combining the upper and lower bounds, the ground state
energy satisfies
\bea
-12 n_h L t \le E_g \le -12 n_h L t + t O(\frac{n_h^2}{L}) + 3V \frac{n_h^2}{L}.
\label{eq:stability}
\eea

So far, we have assumed the same number of holes $n_h$ in each layer.
In fact, the result of Eq. \eqref{eq:stability} can be straightforwardly
generalized to the case with different number of holes in different
layers as
\bea
-4N_h t \le E_g \le -4N_h t +t O (\frac{n^2_{h,m}}{L})
+3V O(\frac{n^2_{h,m}}{L}), \ \ \
\eea
where $N_h$ is the sum of hole numbers of all layers,
and $n_{h,m}$ is the maximal layer hole number.
As a result, the generalized Nagaoka trial state becomes
degenerate with the energy of the ground state $E_g$ in
the thermodynamic limit, when the maximal layer hole number
$n_{h,m}$ in each layer scales as $L^\alpha$ with $\alpha<\frac{1}{2}$.
Hence, the total bulk hole number can increase to the order
of $L^{\frac{3}{2}}$,
which is higher than the single band case  in the 3D cubic
lattice  \cite{Tian1991,Shen1993a}, in which $\alpha<\frac{6}{5}$.
This is due to the combined effect of the quasi-2D band
structure and the Hund's coupling.

\subsection{Instability against a single spin-flip}
\label{sect:SKA}

\begin{figure}
\centering
\epsfig{file=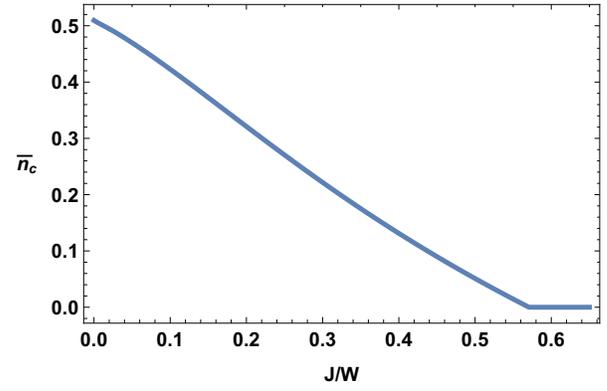,width=0.9\linewidth}
\caption{
The electron density $\bar{n}_c$ below which the generalized Nagaoka state
becomes unstable to a single spin flip. $W = 8t$ is the band width for a 2D square band.
When $J/W$ is larger than a critical value around $0.57$,
a single spin flip is not sufficient to destabilize the state
for any electron density.}
\label{fig:spinflip}
\end{figure}

The Shastry-Krishnamurthy-Anderson method \cite{Shastry1990,fazekas}
provides a useful way to identify a region of instability of the Nagaoka trial state.
Here we generalize it for the multi-orbital systems.
The method considers modifying the state by removing a spin-up particle from the Fermi surface of one orbital layer,
flipping its spin, and adding it back
to the bottom of spin-down band.
The $U = \infty$ limit is enforced by projecting out
doubly occupied orbitals in real space.
This procedure takes the form
\begin{equation}
|\phi_{l,a}\rangle = \prod_{\textbf{r}\in l} [1-n_{a\uparrow}(\textbf{r})n_{a\downarrow}(\textbf{r})] d_{a\downarrow}^\dagger(\textbf{q})d_{a\uparrow}(\textbf{k$_F$})
|\psi_t\rangle,
\end{equation}
where $\textbf{r}$, $\textbf{q}$, and $\textbf{k$_F$}$ are all in layer $l$ and orbital type $a$. Here $\textbf{k$_F$}$ is a Fermi wave vector 
and $\textbf{q}$ is a momentum vector at the band bottom. The energy difference between $|\phi_{l,a}\rangle$ and $|\psi_t\rangle$ as a function of the filling can be computed, and a region of instability can be identified when $|\phi_{l,a}\rangle$ has lower energy.
Since a spin has been flipped in only one band relative to the fully polarized generalized Nagaoka state, the energy difference between $|\phi_{l,a}\rangle$ and $|\psi_t\rangle$ follows only from the kinetic energy change in the up and down spins in layer $l$ and from the Hund's interaction between orbital $a$ and the other two orbital types at sites on layer $l$.

The kinetic energy difference due to a spin flip in a single band on
a square lattice, as evaluated in Ref. [\onlinecite{Shastry1990}] is
\bea
\label{eq:kinchange_2D}
\Delta E_K = -\epsilon_F - \frac{E_{Nag}}{n_h} - 4t\frac{n_h}{L^2}\left[1-\left(\frac{E_{Nag}}{4tn_h}\right)^2\right],
\eea
where $E_{Nag}/L^2 = \int_{-4t}^{\epsilon_F}\epsilon\rho_{2D}(\epsilon)d\epsilon$ and $n_h/L^2 = \int_{\epsilon_F}^{4t}\rho_{2D}(\epsilon)d\epsilon$.
Here $\rho_{2D}(\epsilon) = \frac{1}{2\pi^2t}\Theta(4t-|\epsilon|)K(1-\epsilon^2/16t^2)$, with $K$ a complete elliptic integral of the first kind, is the density of states for a square lattice with nearest neighbor hopping. $E_h$ is the kinetic energy of a single-band Nagaoka state with $n_h$ holes on a square lattice, corresponding to the state in Eq. \eqref{eq:slater}.

Since $H^V$ is only concerned with the number of holes, flipping a single spin does not change $E_V$, i.e., $\Delta E_V=0$,
as can be verified by explicit calculation.
What remains is then to evaluate $\Delta E_J = \langle \phi_{l,a} | H^J | \phi_{l,a} \rangle / \langle \phi_{l,a} | \phi_{l,a} \rangle$.
Expressing the spin operators in terms of $d_\sigma$ and $d^\dagger_\sigma$, the only terms that can possibly contribute to $\langle H^J \rangle$ are those involving a down spin operator in only orbital $a$,
\bea
\Delta E_J=
\frac{J}{2} \sum_{\substack{\textbf{r} \in l,\\ b \neq a}}
\frac{\langle \phi_{l,a}| n_{a\downarrow}(\textbf{r}) n_{b\uparrow}(\textbf{r})|\phi_{l,a} \rangle}{\langle \phi_{l,a}|\phi_{l,a}\rangle}.
\eea
Since $|\phi_{l,a}\rangle$ is a direct product of wavefunctions
of each layer, $n_{a\downarrow}$ and $n_{b\uparrow}$
are uncorrelated, in spite of the strong intra-layer
correlations.
Then we have
\bea
\Delta E_J&=&
\frac{J}{2} \sum_{\substack{\textbf{r} \in l,\\ b \neq a}}
\frac{\langle \phi_{l,a}| n_{a\downarrow}(\textbf{r})
|\phi_{l,a}\rangle \langle \phi_{l,a}|
 n_{b\uparrow}(\textbf{r})|\phi_{l,a} \rangle}{
 |\langle \phi_{l,a}|\phi_{l,a}\rangle|^2}\nn \\
&=& \frac{J}{2} L^2 \sum_{b\neq a} \bar {n}_{la\downarrow}\bar{n}_{lb\uparrow},
\eea
where $\bar {n}_{la\downarrow}$ and
$\bar{n}_{lb\uparrow}$ are independent of $\textbf{r}$
since $|\phi_{l,a}\rangle$ is a momentum eigenstate.
It is easy to evaluate that $\bar{n}_{la\downarrow}=1/L^2$
and $\bar{n}_{lb\uparrow}=1-n_h/L^2$, hence,
\bea
\Delta E_J= J \bar n,
\eea
where $\bar{n} = 1-\frac{n_h}{L^2}$ is the electron density per site in the orbital plane.

Combining the Hund's interaction energy change with the kinetic energy change from Ref. [\onlinecite{Shastry1990}], we have
\bea
&&\Delta E (n)/t=(\Delta E_K +\Delta E_V+\Delta E_J)/t\nn \\
&=& -w -4 (1-\bar n) +
\frac{wy}{1-\bar n} (\frac{wy}{4}- 1)
+ \frac{J}{t}\bar n,
\label{eq:energy_2D}
\eea
where $w=\epsilon_F/t$,
$y=\int^1_{-4/w} x \rho_{2D}(x\epsilon_F) dx$, and
$w(\bar n)$ is determined through the relation of
$\bar n =\int_{-4/w}^1 \rho_{2D}(x\epsilon_F) dx$.
The electron density $\bar n_c$ below which the generalized Nagaoka state
becomes unstable to a single spin flip can be solved by requiring
$\Delta  E(\bar n_c)=0$.
The critical density $\bar{n}_c(J/W)$ is plotted in Fig. \ref{fig:spinflip} where $W = 8t$ is the band width.
As $J$ increases, the ferromagnetic ground state 
becomes more and more stable.
There exist a value of $J/W \approx 0.57$, beyond which the
Nagoka state is stable against a single spin-flip at any
electron density.

\section{The stability of the $t_{2g}$ Nagaoka state
with 3D band structure}
\label{sect:3Dbands}

In this section, we consider the stability of the 3D Nagaoka
state with $t_{2g}$-orbitals and a 3D band structure.

Consider a Hamiltonian
\begin{equation}
H = H^K + H^U + H^V+ H^J
\end{equation}
as before, but where the kinetic terms now allow for perpendicular hopping within the same $d$ orbital. Electrons now hop along the cube, remaining in the same orbital type, and the system is now composed of three cubic orbital bands. Explicitly, the perpendicular $d_{xy}$ orbital hopping modifies $H^K_{xy}$ to
\bea
H^K_{xy} &=& t\sum\limits_{\textbf{r},\sigma} \big(d_{xy,\sigma}^\dagger(\textbf{r})d_{xy,\sigma}(\textbf{r}+\hat{x})
+ d_{xy,\sigma}^\dagger(\textbf{r})d_{xy,\sigma}(\textbf{r}+\hat{y})
\nn \\
&+& d_{xy,\sigma}^\dagger(\textbf{r})d_{xy,\sigma}(\textbf{r}+\hat{z})+ h.c.\big).
\eea
The hopping Hamiltonians of the $d_{yz}$ and $d_{xz}$ orbital
bands can be similarly modified.

We can prove in the case where each orbital band has exactly
one hole that the Nagaoka state is the unique ground state, up to trivial
spin degeneracy.
This can be done through Perron-Frobenius methods used in Refs. [\onlinecite{Tasaki1989}] and [\onlinecite{LiFM2015}] as follows.
Since the off-diagonal matrix elements, the
hopping terms and spin-flipping Hund's coupling terms, all have
negative matrix elements in the basis used in Ref. [\onlinecite{LiFM2015}], the non-positivity condition is satisfied.
Now let us check the transitivity condition.
Within each orbital band, it is satisfied, since any two spins can be exchanged by repeatedly exchanging neighboring spins by cycling the hole around the square plackets.
Spins in different orbitals can be exchanged by moving the spins to the same site, exchanging them using the Hund's coupling, and returning the spins to their original positions following the method presented in Ref. [\onlinecite{LiFM2015}].
Since both the connectivity and non-positivity
conditions of the Perron-Frobenius theorem are satisfied, and the ground state must be a positive-weight superposition of all basis elements.
Since the maximum total-spin state is symmetric under exchange of any two spins, it has nonzero overlap with this positive-weight superposition, and thus the positive-weight superposition must be a maximum total-spin state due to the SU(2) symmetry.

With the Nagaoka-like state established as the ground state when there is a single hole in each of the three orbital bands, the stability of this state can be analyzed in the presence of multiple holes as
in the case of 2D band structure studied above.
Define the reference state $|R_{a,\uparrow}\rangle$ where all momentum states $\textbf{k}$ in the 3D Brillouin zone are filled.
In this case, there is no need for a layer index.
Adding $n_h$ holes to each band then takes the form of Eq. \eqref{eq:slater} with no layer index.
The trial state of interest is then
\bea
	|\psi_t\rangle=|h_{xy,\uparrow}\rangle \otimes |h_{yz,\uparrow}\rangle \otimes |h_{xz,\uparrow}\rangle,
\eea
which corresponds again to filling holes up to their Fermi energy
in each band.

Now let us calculate the energy expectation value of the trial
state $|\psi_t\rangle$.
$E_U = 0$ since no orbital is doubly occupied, and $E_J = 0$,
since the trial state is fully spin-polarized.
In this case,
\bea
E_V &=& V \sum_{r, a > b} \frac{\langle \psi_t |1 - n_a(\textbf{r})|\psi_t \rangle \langle \psi_t| 1 - n_b(\textbf{r}) |\psi_t \rangle}{\langle \psi_t | \psi_t \rangle^2}  \nn \\
&=& 3\frac{n_h^2}{L^3}V,
\eea
since the $n_h$ holes in each band are now distributed over $L^3$ sites.
The kinetic energy $E_K$ can now be evaluated for each band as
\bea
-6n_ht + tL^3 \int_0^{k_0} \frac{k^4dk}{4\pi^2}
= -6n_ht + tO(\frac{n_h^{5/3}}{L^2}),
\eea
where $6L^3/\pi^2 k_0^3 \approx n_h$.
Including all three bands gives a factor of $3$, and the resulting
upper bound for the ground state energy is
\bea
E_g \leq -18n_ht + tO(\frac{n_h^{5/3}}{L^2}) + 3V \frac{n_h^2}{L^3}.
\eea
The lower bound follows similarly to the $t_{2g}$ case.
In the case of 3D bands, the lower bound to the kinetic energy follows from maximizing the number of possible hole hoppings, which allows each hole to hop to six neighboring sites.
Thus, each of the three band contributes $-6n_ht$ to the lower bound, and the ground state energy is bounded by
\bea
-18n_ht \leq E_g \leq -18n_ht + tO(\frac{n_h^{5/3}}{L^2}) + 3V \frac{n_h^2}{L^3}.
\eea
This Nagaoka-type trial state will be degenerate with the ground state in the thermodynamic limit when $n_h$ scales as $L^\alpha$ where
$\alpha < \frac{6}{5}$.

\begin{figure}
\centering
\epsfig{file=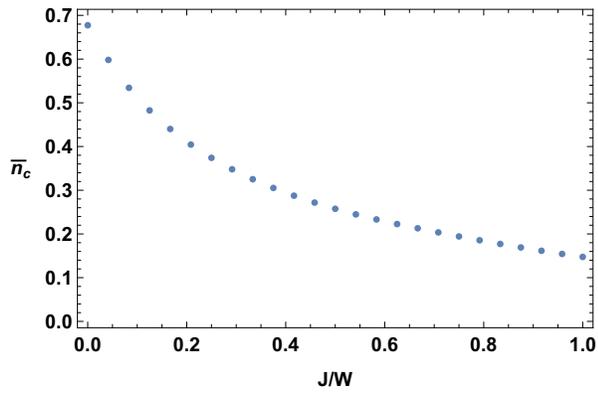,width=0.9\linewidth}
\caption{ The critical value of electron density
$\bar{n}_c$ below which the Nagaoka-like state $|\psi^{3D}_t\rangle$ becomes unstable to a single spin flip.
$W = 12t$ is the band width for the 3D cubic bands.
Unlike in the case of 2D bands, $\bar{n}_c$ does not
approach zero for finite $J/W$ because the 3D density
of states vanishes at low density limit.
}
\label{fig:spinflip3D}
\end{figure}

An instability analysis of the ferromagnetic state for the case with 3D band structure can be performed as in Sec. \ref{sect:SKA}.
The resulting kinetic energy change is \cite{Shastry1990}
\bea
\Delta E_K = -\epsilon_F - \frac{E_{Nag}}{n_h} - 6t\frac{n_h}{L^3}\left[1-\left(\frac{E_{Nag}}{6tn_h}\right)^2\right]
\label{eq:kinchange_3D}
\eea
where $E_{Nag}/L^3 = \int_{-6t}^{\epsilon_F}\epsilon\rho_{3D}(\epsilon)d\epsilon$ and 
$n_h/L^3 = \int_{\epsilon_F}^{6t}\rho_{3D}(\epsilon)d\epsilon$,
with $\rho_{3D}(\epsilon)$ the density of states on a 3D simple cubic lattice.
The energy change due to $H^V$ and $H^J$ can be shown to take the same form as in the case of 2D bands with $\Delta E_V = 0$ and $\Delta E_J = J \bar{n}$ where $\bar n$ is the electron density in each orbital band
defined as $\bar n=1- n_h/L^3$.
Then the total energy change can be expressed as
\bea
\frac{\Delta E(\bar n)}{t} &=&-w-6 (1-\bar n)
+\frac{wy}{1-\bar n}(\frac{wy}{6}-1)+\frac{J}{t}
\bar n,
\nn \\
\label{eq:energy_3D}
\eea
where $w=\epsilon_F/t$, $y=\int^1_{-6/w}x\rho_{3D}(x\epsilon_f)dx$
and $w(\bar n)$ is determined by
$\bar n=\int^1_{-6/w} \rho_{3D}(x\epsilon_F) dx$.

Again the critical density $\bar n_c(J/W)$ below which the Nagaoka-like state becomes unstable to a single spin flip
is solved and shown in Fig. [\ref{fig:spinflip3D}], where $W = 12t$ is the band width.
At $J=0$, the value of $\bar{n}_c=0.68$ is consistent with
previous results in Ref. [\onlinecite{Shastry1990}] for
a model with a single 3D band.
The Hund's coupling further stabilizes the Nagaoka-like
state, which is similar to the case with 2D band
structure in Sect. \ref{sect:SKA}.
However, a significant difference is that $\bar n_c$ does
not drop to zero even at large values of $J/W$.

The different behavior of $\bar n_c$ for 2D and 3D bands is due to
the different scalings of density of states at low energy.
It is easy to check that in the low density limit, the energy
costs of a single spin-flip in Eq. \eqref{eq:energy_2D} and Eq. \eqref{eq:energy_3D} can be expanded to the leading order as
\bea
\Delta E_d(\bar n_d) \approx -(\epsilon_F-\epsilon_b) + J \bar n_d,
\label{eq:cost}
\eea
where 
$\epsilon_b$ is the band bottom energy
and $\bar n_d$ is the particle density for $d$-dimensional bands.
In the low density limit, $\epsilon_F-\epsilon_b \propto (\bar n_d)^{2/d}$.
In 3D, the kinetic energy change in Eq. \eqref{eq:cost} dominates 
the Hund's coupling energy cost, allowing a single spin-flip to lower the total energy and destablize the Nagaoka state.
By contrast, both terms in Eq. \eqref{eq:cost} scale the same in 2D,
and a single spin-flip costs energy when $J$ is large enough.

\section{Conclusions and Discussions}
\label{sect:conclusion}

We have studied the stability of the generalized Nagaoka ferromagnetic state in a 3D cubic lattice with the $t_{2g}$-orbitals.
Applying the bounding method of Refs. [\onlinecite{Tian1991,Shen1993a}], for a cubic lattice
with size $L\times L\times L$ and a quasi 2D $t_{2g}$-orbital
band structure, the fully polarized Nagaoka state becomes
degenerate with the ground state as $L\to \infty$ when the
number of holes in each orbital plane scales slower than $L^{\frac{1}{2}}$, or, the total hole number scales slower than  $L^{\frac{3}{2}}$.
For the case with 3D band structure, we have generalized the
Nagaoka theorem to the case that each orbital has a single hole.
Again for the multi-hole case, the fully polarized Nagaoka ferromagnetic state remains degenerate with the ground state at $L\to \infty$
when the hole number scales slower than $L^{\frac{6}{5}}$.
These results apply in the limit of $U\to \infty$ and
arbitrary ferromagnetic Hund's coupling $J>0$ and
inter-orbital repulsion $V\ge 0$.
We have also examined the stability of the orbital-generalized
Nagaoka states against a single spin flip for both
quasi-2D and 3D band structures.
In both cases, the instability region shrinks as
the Hund's coupling increases.

The above bounding estimation only proves the degeneracy
of the Nagaoka-type ferromagnetic state with the
ground state but does not prove the uniqueness of the ground state.
Hence, even within the above bounds, the above results actually
do not prove the ground state ferromagnetism for the multi-hole case.
Nevertheless, the stability of the ground state ferromagnetism
is still conceivable.
The above analysis does not imply that the fully polarized
state must break down when the hole number exceeds
the above bounds.
Recent numerical calculations based on the
density-matrix-renormalization-group method
have shown evidence of the stability of Nagaoka
state at finite hole densities for the 2D single-band case
\cite{Liuyao2012}, although exact proof remains
an open question.

The above study is not just of academic interest.
In fact, itinerant ferromagnetism has been discovered in the $t_{2g}$-orbital active materials SrRuO$_3$, which is
a weak ferromagnet with partial polarization and its
Curie temperature $T_c\approx 160K$ \cite{Jeong2013,Koster2012,Cao1997}.
Its electronic structure can be modeled by the multi-orbital
Hubbard model with the quasi-2D band structure and
the prominent Hund's coupling.
Certainly, the filling is 4/3 electron per orbital on each
site, and thus significantly way from the half-filling
which corresponds to one electron per orbital.
The real system of SrRuO$_3$ implies that the
Hund's rule-facilitated itinerant ferromagnetism
may remain stable at finite values of $U$ and away
from half-filling.
Our work provides a useful guidance for studying itinerant
ferromagnetism in this class of materials.

\acknowledgments
E. B. and Y. L. are supported by the U.S. Department of Energy, Office of Basic Science, Division of Materials Sciences and Engineering, Grant No. DE-FG02-08ER46544.


\appendix*
\section{Stability of One-Band Nagaoka State}
\label{appdx:Tian}

The method of applying the Gershgorin circle theorem to study the stability of the Nagaoka state was used for a one-band Hubbard model on a square lattice in Ref. [\onlinecite{Tian1991}], where the model used corresponds to $H = H^U + H^K_{xy}$. The method, as generalized to a $d$-dimensional hypercubic lattice in Ref. [\onlinecite{Shen1993a}], is reviewed here for completeness. For a single-band Hubbard model, the fully polarized Slater determinant state $|h\rangle$ in Eq. \eqref{eq:slater} can be taken as a variational trial state. In $2D$ bands on a simple square lattice, this trial state corresponds to a single orbital plane $a, l$. The trial state can be generalized to $d$-dimensional bands on a hypercubic lattice by taking $\br$ and $\textbf{k}$  to be $d$-dimensional lattice and momentum vectors, respectively.

The energy $E_K$ of this trial state was evaluated by considering the dispersion for a $d$-dimensional hypercubic lattice. In terms of the hole number occupation picture,
\bea
H^K = -2dt\sum_{\mathbf{k},\sigma}\left(1-\frac{\mathbf{k}^2}{2d}\right)(1-n_\sigma(\mathbf{k})).
\eea

This parabolic dispersion holds in the limit $n_h/L^d \to 0$. The energy of the trial state can then be evaluated as
\bea
E_K = -2dtn_h + tL^d\int_0^{k_0} \frac{\Omega_d k^{d+1}dk}{(2\pi)^d}
\eea
where $\Omega_d$ is the surface area of a unit $d$-sphere and $(L/2\pi)^d V_d k_0^d \approx n_h$, with $V_d$ the volume of a unit $d$-sphere. The upper bound on the ground state energy is then

\bea
E_g \leq -2dtn_h + tO\left(\frac{n_h^{\frac{d+2}{d}}}{L^2}\right).
\eea

The lower bound follows from the Gershgorin circle theorem, which states that for any eigenvalue $\lambda$ of a square matrix $H$, there exists a row $i$ such that
\begin{equation}
|\lambda - H_{ii}| \leq \sum_{j \neq i} |H_{ij}|.
\end{equation}
It follows that a lower bound on the ground state energy is given by
\begin{equation}
	\min_i\{H_{ii} - \sum_{j \neq i}|H_{ij}|\} \leq E_g.
\end{equation}

Intuitively, the lower bound is the configuration that minimizes the energy of an analogous bosonic system, where sign changes from hopping are neglected. For the 2D square lattice, this configuration corresponds to placing all holes on either the even or odd sublattice, allowing each hole to hop to four neighboring sites. Thus,
\begin{equation}
	-4n_ht \leq E_g.
\end{equation}
The lower and upper bounds on the ground state energy coincide in the thermodynamic limit as long as $n_h \sim L^{\alpha}$ where $0 \leq \alpha < \frac{2d}{d+2}$. Thus, the fully spin-polarized trial state is a ground state that remains stable for a number of holes that grows sufficiently slowly.

%

\end{document}